\documentclass[aps,prl,twocolumn,amsmath,amssymbs,longbibliography, superscriptaddress]{revtex4-1}

\usepackage{graphicx}
\usepackage{amsmath,amsfonts,amssymb}
\usepackage{epsfig}
\usepackage{wrapfig}
\usepackage{bbm}
\usepackage[usenames]{color}
\usepackage{array}
\usepackage{times}
\usepackage{epstopdf}

\newcommand{\be}{\begin{equation}}
\newcommand{\ee}{\end{equation}}

\def\red{\color{black}}

\def\black{\color{black}}

\def\W{\omega}

\begin{document}


\title{Interference of Temporally Distinguishable Photons Using Frequency-Resolved Detection}

\author{Venkata Vikram Orre}
\affiliation{Joint Quantum Institute,NIST/University of Maryland, College Park, Maryland 20742 USA}
\affiliation{Department of Electrical and Computer Engineering and IREAP, University of Maryland, College Park, Maryland 20742, USA}

\author{Elizabeth A. Goldschmidt}
\affiliation{Joint Quantum Institute,NIST/University of Maryland, College Park, Maryland 20742 USA}
\affiliation{U.S. Army Research Laboratory, Adelphi, Maryland 20783, USA}

\author{Abhinav~Deshpande}
\affiliation{Joint Quantum Institute,NIST/University of Maryland, College Park, Maryland 20742 USA}
\affiliation{Joint Center for Quantum Information and Computer Science, NIST/University of Maryland, College Park, Maryland 20742, USA}

\author{Alexey~V.~Gorshkov}
\affiliation{Joint Quantum Institute,NIST/University of Maryland, College Park, Maryland 20742 USA}
\affiliation{Joint Center for Quantum Information and Computer Science, NIST/University of Maryland, College Park, Maryland 20742, USA}

\author{Vincenzo Tamma}
\affiliation{School of Mathematics and Physics, University of Portsmouth, Portsmouth PO1 3QL, United Kingdom}
\affiliation{Institute of Cosmology and Gravitation, University of Portsmouth, Portsmouth PO1 3FX, United Kingdom}

\author{Mohammad Hafezi}
\affiliation{Joint Quantum Institute,NIST/University of Maryland, College Park, Maryland 20742 USA}
\affiliation{Department of Electrical and Computer Engineering and IREAP, University of Maryland, College Park, Maryland 20742, USA}
\affiliation{Department of Physics, University of Maryland, College Park, Maryland 20742, USA}

\author{Sunil Mittal}
\email[Email: ]{mittals@umd.edu}
\affiliation{Joint Quantum Institute,NIST/University of Maryland, College Park, Maryland 20742 USA}
\affiliation{Department of Electrical and Computer Engineering and IREAP, University of Maryland, College Park, Maryland 20742, USA}

\begin{abstract}
We demonstrate quantum interference of three photons that are distinguishable in time, by resolving them in the conjugate parameter, frequency.  We show that the multiphoton interference pattern in our setup can be manipulated by tuning the relative delays between the photons, without the need for reconfiguring the optical network. Furthermore, we observe that the symmetries of our optical network and the spectral amplitude of the input photons are manifested in the interference pattern. We also demonstrate time-reversed Hong-Ou-Mandel-like interference in the spectral correlations using time-bin entangled photon pairs. By adding a time-varying dispersion using a phase modulator, our setup can be used to realize dynamically reconfigurable and scalable boson sampling in the time domain as well as frequency-resolved multiboson correlation sampling.
\end{abstract}

\maketitle


The nonclassical interference of two or more photons in an optical network is the fundamental phenomenon enabling many algorithms used in linear optics quantum computing \cite{Knill2001, Carolan2015, Ladd2010, Pan2012}, quantum communications \cite{Bouwmeester1997, Mattle1996, Lo2014}, metrology \cite{Dowling2008,Lemos2014} and boson sampling \cite{Motes2014, Aaronson2011, Spagnolo2013b}. Quantum interference, such as, Hong-Ou-Mandel (HOM) and Shih-Alley interference \cite{Hong1987,Shih1988}, usually requires photons that are identical in their temporal and spectral degrees of freedom. Any distinguishability in the photons at the detectors leads to a reduction in the interference. The difficulty in experimentally generating identical photons has prompted strong interest in developing “real world” optical networks enabling the interference of nonidentical photons \cite{Legero2003, Legero2004}. Recently, it was shown that nonclassical interference can be observed between photons completely distinguishable in time or frequency by exploiting correlation measurements in the corresponding conjugate parameter \cite{Zhao2014, Vittorini2014, Tamma2015, Laibacher2018, Wang2018}. Remarkably, the interference can occur for any values of the input frequencies (or times) as long as the detector resolution in the conjugate parameter is sufficient to make the detectors `blind' to the spectral (or temporal) distinguishability of the photons. Furthermore, the temporal or spectral distinguishability can actually be used as a resource, for example, to reveal spectral properties of the input photons and the symmetries of the optical network \cite{Tamma2015,Laibacher2018}.

Many experiments have demonstrated interference of two photons that are distinguishable in frequency or time by resolving them in the conjugate parameter \cite{Zhao2014,Vittorini2014,Gerrits2015,Jin2015}. Scaling these spectrally or temporally resolved interference phenomena to a larger number of photons can enable, for example, multiboson correlation sampling experiments where sampling over temporal or spectral modes, in addition to spatial modes, can relax the requirements on generating identical photons and could demonstrate quantum supremacy \cite{Laibacher2015, Tamma2016b, Tamma2016, Laibacher2018b}. Indeed, time-resolved interference of three photons with different frequencies was demonstrated very recently where the temporal correlations between detected photons were manipulated using a spatial network of beam splitters \cite{Wang2018}. In contrast, the complementary phenomenon, that is, frequency-resolved interference of multiple photons that are separated in time, allows for the convenient manipulations of spectral correlations by tuning the relative delays between photons, without reconfiguring the spatial network \cite{Laibacher2018, Laibacher2018b}. This scheme can operate in a single spatial (transverse) mode and, therefore, enable the realization of scalable temporal boson sampling using time-varying dispersion \cite{Pant2016}. However, such frequency-resolved interference of more than two photons has not yet been demonstrated.

Here, we demonstrate frequency-resolved quantum interference of three photons that are completely distinguishable in time. We show that the interference observed in the spectral correlations of detected photons can be manipulated by changing the relative delays between the photons at the input. The interference is completely wiped out for longer delays between photons, that is when the spectral resolution of the detectors is not sufficient to erase the temporal distinguishability of the photons. Moreover, we observe that the symmetries of the optical network, and the spectral wave functions of photons are reflected in the measured spectral correlations. Finally, we also demonstrate spectral correlations in inverse-HOM interference using two time-bin entangled photons where both photons arrive either ``early'' or ``late'' \cite{Chen2007}. In this case, we observe that the interference is sensitive to the phase between the two components of the entangled state, unlike the case of unentangled photons where interference is insensitive to small fluctuations in delay between the photons. Our experimental setup could easily be extended to introduce time-varying dispersion using phase modulators and realize the temporal boson sampling scheme of Ref.~\cite{Pant2016}.

\begin{figure}
\includegraphics[width=0.48\textwidth]{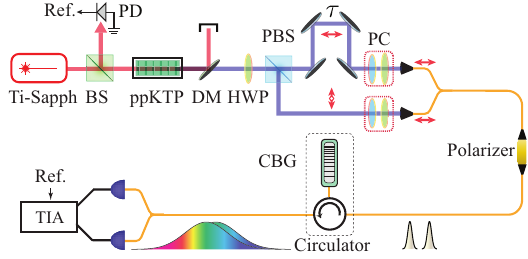}
\caption{\label{fig:1} Schematic of the experimental setup to observe frequency-resolved two-photon interference. Orthogonally polarized photon pairs are generated using Type-II SPDC in a PPKTP crystal, separated using a polarization beam splitter (PBS), delayed and recombined using a beam splitter (BS) after polarization (red arrows) rotation in one of the arms. \red A chirped Bragg grating (CBG) with a spectral bandwidth more than that of the generated photon pairs, and a time-interval analyzer (TIA) implement a time-of-flight spectrometer (see the supplemental material). \black HWP: half-wave plate (to generate entangled photon pairs), \red DM: dichroic mirror, \black  PC: polarization controller, PD: photo-diode.
}
\end{figure}

To demonstrate our scheme, we discuss first the interference of two temporally distinguishable photons in our setup (Fig.~\ref{fig:1}). The two-photon interference can be analyzed using the spectral correlation function $\Gamma\left(\W_{1},\W_{2}, \tau\right)$, which is the probability of detecting two photons at the two detectors, with frequencies $\W_{1}$ and $\W_{2}$, respectively, and $\tau = t_{2} - t_{1}$ is the relative delay between photons at the input. The correlation function in our setup is given by \cite{Jin2015, Gerrits2015}
\begin{eqnarray}\label{Gamma_TP}
\nonumber \Gamma\left(\W_{1},\W_{2},\tau \right)  =  \Bigl| &\psi_{1}& \left(\W_{1}\right) \psi_{2} \left(\W_{2}\right) e^{-i\left(\W_{1} t_{1} + \W_{2} t_{2} \right)} \\
                                        +  ~~&\psi_{1}& \left(\W_{2}\right) \psi_{2} \left(\W_{1}\right) e^{-i\left(\W_{2} t_{1} + \W_{1} t_{2} \right)} \Bigr|^{2},
\end{eqnarray}
where $\psi_{1(2)} \left( \W \right)$ is the spectral wavefunction of the first (second) photon. The spectral correlation function depends on $t_{1}$ and $t_{2}$ only through the relative delay $\tau$ and exhibits interference fringes as a function of $\left(\W_{1}-\W_{2}\right)$, with fringe separation $2\pi/\tau$ \cite{Pant2016}. Furthermore, because the photons are in a single spatial mode, the unitary transformation describing our optical network adds an overall phase to the photonic wave functions, and therefore, does not contribute to the interference.

\begin{figure}[!ht]
 \centering
\includegraphics[width=0.48\textwidth]{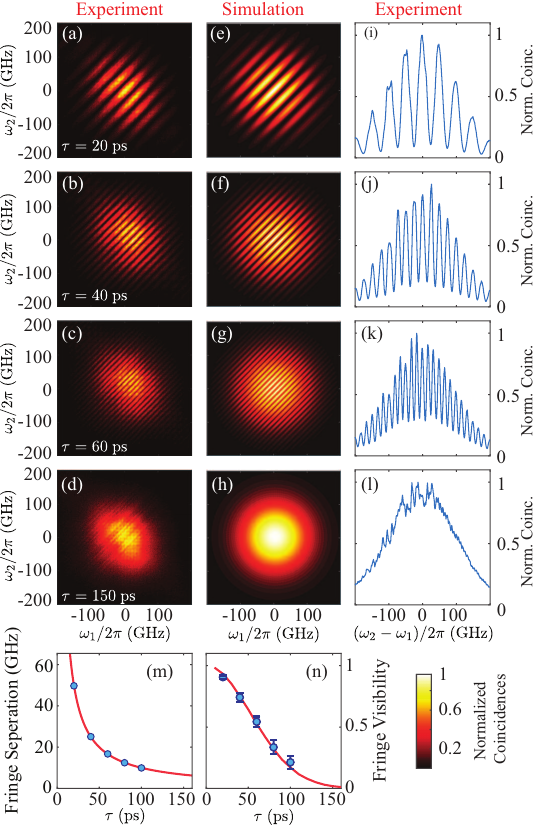}
\caption{\label{fig:2}(a)-(d) Measured and, (e)-(h) simulated spectral correlations $\left[\Gamma\left(\W_{1},\W_{2},\tau \right) \right]$ between two photons with varying relative delay $\tau$ at the input. (i)-(l) Measured coincidence counts (maximum normalized to unity) as a function of frequency separation $\left(\W_2-\W_1\right)$. Measured (blue dots) and simulated (red line) (m) fringe separation, and (n) visibility as function of the delay $\tau$. \red  The delay between the photons was calibrated using HOM interference. \black}
\end{figure}

In our experiment, we generate photon pairs using spontaneous parametric down-conversion (SPDC) (Fig.~\ref{fig:1}). A periodically poled potassium titanyl phosphate (PPKTP) crystal (30 mm length) is pumped using a pulsed ($\approx 1.6 ~\text{ps}$) Ti:sapphire laser \red($\approx 775.5~\text{nm}, 50~\text{mW}$) \black which generates orthogonally polarized, spectrally degenerate photon pairs at telecom wavelengths via a Type-II collinear SPDC. We separate the two orthogonally polarized photons using a polarization beam splitter (PBS) and introduce a relative delay $\left(\tau \right)$ between them. We rotate the polarization in one of the arms such that the two photons are identically polarized and collect them into a single fiber using a beam splitter. We then use a chirped fiber Bragg grating (CBG), two superconducting nanowire detectors (SNSPDs), and a time-interval analyzer (TIA) to measure the spectral correlations between photons. This setup realizes a time-of-flight spectrometer where the arrival time of dispersed photons is used to infer their frequency spectrum \cite{Avenhaus2009,Gerrits2015,Jin2015,Mittal2017,Davis2017}. Specifically, the frequency $\W_{i}$ of a photon detected at the detector $i$ is related to the time-of-arrival $t^{d}_{i}$ at the detector as $\left(\W_{i} - \W_{0} \right) = \left(t^{d}_{i} - t^{d}_{0i}\right)/\phi^{\prime\prime}$. Here, $\W_{0}$ is the peak frequency of the photonic spectral wave packet and $t^{d}_{0i}$ is the peak arrival time of the photonic temporal wave packet at the detector $i$. $\phi^{\prime\prime} \simeq 3196 ~\text{ps}^{2}$ is the group delay dispersion (GDD) of the CBG. In our measurements, we set the central frequency $\W_{0}$ (corresponding to the time $t^{d}_{0}$) to be zero such that $\W_{i}$ is actually the detuning from the central frequency. The spectral resolution $\left(\delta\W \right)$ of our spectrometer is limited by the timing jitter ($\approx$ 100 ps) of the nanowire detectors and is $\approx 5~\text{GHz}$. Furthermore, the finite delay between the input photons contributes to the timing uncertainty in $t^{d}_{0}$ and marginally lowers the spectral resolution of our spectrometer for input delay values approaching the inherent timing jitter of the detectors.

Figures~\ref{fig:2}(a)-\ref{fig:2}(d) show quantum interference fringes in the measured spectral correlations $\left[\Gamma\left(\W_{1},\W_{2} \right), ~\text{twofold coincidences} \right]$ for different delays between the two photons. The interference fringes can be seen more clearly by plotting the number of coincidences as a function of the frequency separation $\W_{2}-\W_{1}$ [Figs.~\ref{fig:2}(i)-\ref{fig:2}(l)]. As expected, the fringe separation decreases as $1/\tau$ [Fig.~\ref{fig:2}(m)]. Moreover, we see that the visibility (see the supplemental material) of the interference decreases with increasing delay [Fig.~\ref{fig:2}(n)], disappearing completely for $\tau \gtrsim 150 ~\text{ps}$ [Fig.~\ref{fig:2}(d)]. This is because of the residual distinguishability following spectrally resolved detection for time delays that approach the inverse of the spectral resolution $\delta\W$. The interference visibility could, in principle, be restored by increasing the spectral resolution of the detector so that the condition $\delta\W \gg 1/\tau$ is satisfied \cite{Laibacher2018,Laibacher2018b, Pant2016}. Our experimental results agree well with the simulation results [Figs.~\ref{fig:2}(e)-\ref{fig:2}(h)]. We note that similar interference in spectral correlations has been observed in Refs.~\cite{Gerrits2015, Jin2015} using the two spatial modes of a HOM interference setup. By contrast, in our setup, the two delayed photons are in a single spatial mode. Furthermore, the large GDD of our CBG allows us to observe interference between photons that are separated by delays as long as $100~\text{ps}$, which is more than $50$ times the single-photon temporal pulse widths (estimated to be $\approx 1.55~\text{ps}$).

\begin{figure}
 \centering
\includegraphics[width=0.48\textwidth]{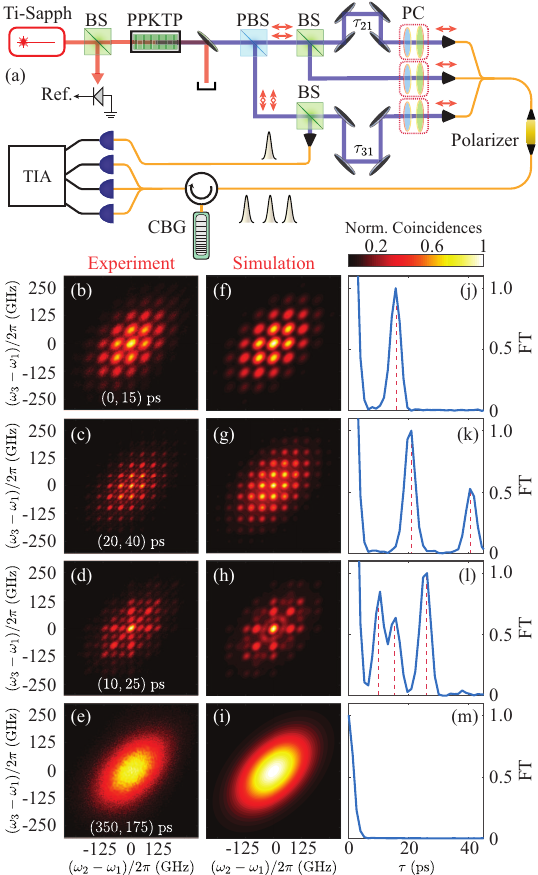}
\caption{\label{fig:3}(a) Schematic of the setup to observe frequency-resolved three-photon interference. The PPKTP crystal is strongly pumped to efficiently generate two pairs of photons which are then probabilistically separated using a PBS and two nonpolarizing beam splitters. (b)-(e) Measured and, (f)-(i) simulated spectral correlations between three photons (heralded by the fourth photon) with relative delays $\left(\tau_{21}, \tau_{31} \right) \approx \left(0, 15 \right); \left(20, 40 \right);  \left(10, 25 \right); \left(350, 175 \right) \text{ps}$. (j)-(m) Fourier transform (FT) of the three-photon correlation function [in (b)-(e)], integrated over $\W_{3}$. The peaks (highlighted by red dashes) indicate beat notes associated with multiple pairwise interferences.}
\end{figure}

Next we discuss the experimental setup and our observation of three-photon interference using frequency-resolved detection. We pump the PPKTP crystal at higher power \red (400 mW) \black to ensure a higher probability of generating two pairs of photons (see the supplemental material). We use a PBS and two nonpolarizing beam splitters to probabilistically split the four photons into four spatial modes [see Fig.~\ref{fig:3}(a)]. We then introduce relative delays between the photons \red using two-photon interference measurements as a calibration tool, \black and combine three of the four spatial modes into a single fiber using a tritter ($3\times 3~$ beam splitter). As before, the three photons are then dispersed using the CBG, separated using a tritter and their spectral correlations are measured using three detectors connected to the TIA. \red The fourth photon is used to trigger the TIA. \black

Figures~\ref{fig:3}(b)-\ref{fig:3}(e) show the measured spectral correlations between the three photons (using fourfold coincidence detections) as a function of the frequency detunings measured at the second and the third detectors, relative to that of the first detector, that is, $\left(\W_{2}-\W_{1}\right)~\text{and} ~\left(\W_{3}-\W_{1}\right)$. We analyze three different delay scenarios (1) $\left(\tau_{21}, \tau_{31} \right) \approx \left(0, 15 \right)  \text{ps}$, (2) $\left(20, 40 \right) \text{ps}$ and (3) $\left(10, 25 \right)  \text{ps}$, where $\left(\tau_{21}, \tau_{31} \right)$ are the delays of the second and the third photon, respectively, relative to the first photon. We observe that the interference landscape changes significantly with the relative delays between photons. For delays symmetric under exchange of two of the photons $\left(\tau_{21}, \tau_{31} \right) \approx \left(0, 15 \right)$ and $\left(20, 40 \right) \text{ps}$, the interference fringes are periodic along both axes. By contrast, in the case of asymmetric delays $\left(10,25\right) \text{ps}$, the constructive correlations are more prominent along the cross sections $\W_{2} = \W_{1}$ (vertical),   $\W_{3} = \W_{1}$ (horizontal), and $\W_{2} = \W_{3}$ (diagonal). However, irrespective of the delays between the photons, we always observe a constructive interference for zero frequency detuning, that is, when $\W_{2} - \W_{1} = 0 = \W_{3} - \W_{1}$ [see Eq.\ref{Gamma_TP}]. We also analyze the scenario when the spectral resolution of our setup is not high enough to erase the temporal distinguishability of photons [Fig.~\ref{fig:3}(e)], and, as expected, we do not observe any interference.

We see that the interference patterns shown in Figs.~\ref{fig:3}(b)-\ref{fig:3}(e) are symmetric under any permutation of the frequency detunings, for instance, $\W_{2} \leftrightarrow \W_{3}$ or $\W_{3} \leftrightarrow \W_{1}$, etc. This permutation symmetry is simply a manifestation of the symmetry of our optical network \cite{Laibacher2018}. We again emphasize that the interfering photons propagate in a single fiber. The tritter, together with the three detectors, at the output simply emulates a number-resolving detector and does not contribute to the interference. Moreover, the interference landscape is also symmetric under reflections $\W_{i} \leftrightarrow -\W_{i}$, for all $i$, where $i$ = 1,2,3 is the detector number, because of the symmetric frequency spectra of the input photons \cite{Laibacher2018}.

The measured three-photon interference is dictated by the $3!$ three-photon detection amplitudes associated with the possible ways in which the three photons can trigger the three detectors. However, it is instructive to integrate the three-photon correlation function over one of the frequencies (here $\W_{3}$) and analyze the reduced interference as a function of the relative frequency detunings at the other two detectors $\left( \W_{2}-\W_{1} \right)$ (see the supplemental material). Fourier analysis of this 1D plot then reveals the beat notes corresponding to the multiple pairwise interference terms between the three photons [Figs.~\ref{fig:3}(j)-\ref{fig:3}(m)]. When the input delay values are configured to be $\left(\tau_{21}, \tau_{31} \right) \approx \left(0, 15 \right) \text{ps}$, there is only one possible delay combination between any two photon pairs and the corresponding Fourier transform shows a single peak (highlighted by the dashed red line) at 15.7 ps. For $\left(20, 40 \right) \text{ps}$, there are two possible combinations and accordingly we observe two peaks in the Fourier transform, at 20.9 ps and 40.4 ps. For $\left(10, 25 \right) \text{ps}$, there are three possible combinations and, as expected, we see three beat notes in the Fourier transform, at 10.4 ps, 15.7 ps, and 26.1 ps. \red The peak delay values agree well with the expected values to within 1.3 ps, the temporal resolution of the Fourier transform. \black

We note that the beam splitters used in our setup to separate photon pairs of the same polarization are not deterministic and lead to possibilities where two photons always arrive with zero delay (see the supplemental material). However, these possibilities do not add any new beat notes to the interference pattern, and could easily be removed using two PPKTP crystals to generate two photon pairs. Nevertheless, our experimental observations match very well with our simulations. We achieve a fidelity of $\approx$0.95 for each of the three scenarios presented in Figs.~\ref{fig:3}(b)-\ref{fig:3}(d). The small loss in observed fidelities is because of the small ellipticity in the joint spectral intensity of the photons (see the supplemental material).

Finally, we demonstrate frequency-resolved interference of two photons which are entangled in their arrival times. In particular, we consider time-bin entangled states of the form $\left|\Psi \right> = \left| 2\right>_{e} \left| 0\right>_{l} -  e^{-i\varphi} \left| 0\right>_{e}  \left| 2\right>_{l}$, where both photons at the input are in the `early' time bin $\left( \text{at time}~t_{1}\right)$ or in the `late' time bin $\left(\text{at}~t_{2} \right)$ and $\varphi$ is the phase associated with the delay between the photons. The spectral correlation function at the output of our optical network (Fig.~\ref{fig:1}) is then given as
\begin{eqnarray}
\nonumber \Gamma\left(\W_{1},\W_{2},\tau \right) = \Bigl| &\psi_{1}& \left(\W_{1}\right) \psi_{2} \left(\W_{2}\right) e^{-i\left(\W_{1}+\W_{2}\right)t_{1}} \\
                                                 +      ~~&\psi_{1}& \left(\W_{2}\right) \psi_{2} \left(\W_{1}\right) e^{-i\left(\W_{1}+\W_{2}\right)t_{2}} \Bigr|^{2}.
\end{eqnarray}
The correlation function now exhibits interference fringes as a function of the two-photon phase $\varphi = \left(\W_{1}+\W_{2}\right)\tau$, where $\tau = t_{2} - t_{1}$ is the relative delay between the photons. This interference is similar to the time-reversed HOM interference where the two photons are path entangled, that is, they arrive together at either port of the beam splitter \cite{Chen2007}. The two photons can then exit the beam splitter in the same port or in different ports, depending on the phase $\varphi$.

\begin{figure}
 \centering
\includegraphics[width=0.48\textwidth]{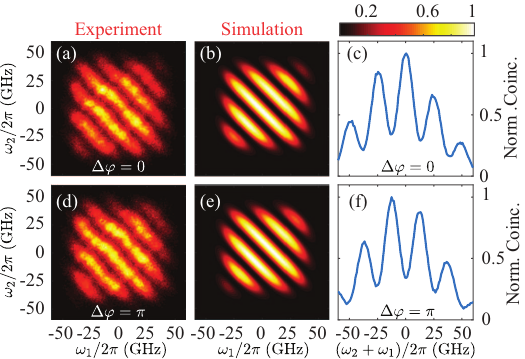}
\caption{\label{fig:4} Measured and simulated spectral correlations for time-bin entangled photon pairs, with the phase factor (a)-(b) $\Delta\varphi = 0$ and, (d)-(e) $\Delta\varphi = ~\pi$.
(c),(f) Measured coincidences as a function of $\left(\W_{1}+\W_{2}\right)$.}
\end{figure}

To generate time-bin entangled photon pairs we add a half-wave plate, set at an angle of $22.5^\circ$, before the PBS in the setup of Fig.~\ref{fig:1}. The HWP acts as a $50:50$ BS in the polarization domain and, when the two-photon spectral wave function is symmetric, it leads to a polarization entangled two-photon state of the form $\left|\Psi\right> = \left|2 \right>_{H} \left|0 \right>_{V} - \left|0 \right>_{H} \left|2 \right>_{V}$ \cite{Kuo2016, Slussarenko2017, Mittal2017}. As before, we use a PBS to introduce a relative delay between the two orthogonal polarization modes and achieve the time-bin entangled state $\left|\Psi \right> = \left| 2\right>_{e} \left| 0\right>_{l} -  e^{-i\varphi} \left| 0\right>_{e}  \left| 2\right>_{l}$. \red The phase $\varphi$ was actively stabilized using a continuous-wave telecom laser with tunable wavelength (see the supplemental material). We also used a bandpass filter \red (with 75 GHz bandwidth) \black to ensure that the two-photon spectral amplitude is symmetric, and verify the entanglement using \red polarization- and time-resolved coincidence measurements (see supplemental material). \black

Figures~\ref{fig:4}(a), and ~\ref{fig:4}(b) show the measured and simulated spectral correlations for the time-bin entangled two-photon state where we set $t_{2} - t_{1} = 40~\text{ps}$, much longer than the single-photon pulsewidths (estimated to be $\approx$ 5.4 ps following the bandpass filter). We observe interference fringes in the correlations as a function of the two-photon phase $\left(\W_{1}+\W_{2}\right)\tau$ [Fig.~\ref{fig:4}(c)]. This contrasts with the interference for a separable state of two-photons with a delay (Fig.~\ref{fig:2}), where interference pattern is rotated by $90^\circ$ because of its dependence on $\left(\W_{1}-\W_{2}\right)\tau$. Furthermore, as in the time-reversed HOM interference, the interference observed here is sensitive to small changes in the two-photon phase $\varphi$. For example, by introducing an additional small delay $\Delta\tau$ (few fs) such that $\Delta\varphi = \left(\W_{1}+\W_{2}\right)\Delta\tau = \pi$, we observe the complimentary interference where the peaks are replaced by troughs and vice versa [Figs.~\ref{fig:4}(d)-\ref{fig:4}(f)]. We note that the marginal decrease in the observed visibility for interference of time-bin entangled photons compared to the unentangled photons (Fig.~\ref{fig:2}) is because of the imperfections in the entangled state preparation and the sensitivity to residual path length fluctuations in the interferometer (see the supplemental material).

In summary, we have demonstrated frequency-resolved interference of three photons that are separated in time using a single dispersive element. Using a larger number of photons and a time-varying dispersion element, such as a phase modulator, our setup could realize temporal boson sampling in a single spatial mode with easily reconfigurable unitary transformation and explore phase transitions in the complexity of sampling \cite{Deshpande2018,Pant2016} (see the supplemental material). Our scheme can also be used to implement scalable multiboson correlation sampling where the photonic correlations are sampled over spatial as well as temporal or spectral modes at the input or output of a random linear optical network with multiple spatial modes \cite{Laibacher2015, Laibacher2018b}. Finally, these experimental results may pave the way to new techniques for the experimental characterization of optical networks and their input photonic states with potential application in quantum information processing and metrology \cite{Oren2017,Laibacher2018,Wang2019}.

\begin{acknowledgments}
This research was supported by the Air Force Office of Scientific Research Multidisciplinary University Research Initiative (AFOSR-MURI Grant No. FA9550-16-1-0323) and the Physics Frontier Center at the Joint Quantum Institute. A.D. and A.V.G. acknowledge funding by ARL CDQI, ARO MURI, NSF PFC at JQI, AFOSR, NSF PFCQC program, DoE ASCR Quantum Testbed Pathfinder program (Award No. DE-SC0019040), and DOE BES QIS program (Award No. DE-SC0019449). V.T. was partially supported by the Office of Naval Research Global (N62909-18-1-2153). We thank Hana Warner and Jessica Christian for help with the experimental setup, and Qudsia Quraishi for the nanowire detectors.
\end{acknowledgments}


\begin{thebibliography}{37}
\expandafter\ifx\csname natexlab\endcsname\relax\def\natexlab#1{#1}\fi
\expandafter\ifx\csname bibnamefont\endcsname\relax
  \def\bibnamefont#1{#1}\fi
\expandafter\ifx\csname bibfnamefont\endcsname\relax
  \def\bibfnamefont#1{#1}\fi
\expandafter\ifx\csname citenamefont\endcsname\relax
  \def\citenamefont#1{#1}\fi
\expandafter\ifx\csname url\endcsname\relax
  \def\url#1{\texttt{#1}}\fi
\expandafter\ifx\csname urlprefix\endcsname\relax\def\urlprefix{URL }\fi
\providecommand{\bibinfo}[2]{#2}
\providecommand{\eprint}[2][]{\url{#2}}

\bibitem[{\citenamefont{Knill et~al.}(2001)\citenamefont{Knill, Laflamme, and
  Milburn}}]{Knill2001}
\bibinfo{author}{\bibfnamefont{E.}~\bibnamefont{Knill}},
  \bibinfo{author}{\bibfnamefont{R.}~\bibnamefont{Laflamme}}, \bibnamefont{and}
  \bibinfo{author}{\bibfnamefont{G.~J.} \bibnamefont{Milburn}},
  \bibinfo{journal}{Nature} \textbf{\bibinfo{volume}{409}}, \bibinfo{pages}{46}
  (\bibinfo{year}{2001}).

\bibitem[{\citenamefont{Carolan et~al.}(2015)\citenamefont{Carolan, Harrold,
  Sparrow, Mart{\'\i}n-L{\'o}pez, Russell, Silverstone, Shadbolt, Matsuda,
  Oguma, Itoh et~al.}}]{Carolan2015}
\bibinfo{author}{\bibfnamefont{J.}~\bibnamefont{Carolan}},
  \bibinfo{author}{\bibfnamefont{C.}~\bibnamefont{Harrold}},
  \bibinfo{author}{\bibfnamefont{C.}~\bibnamefont{Sparrow}},
  \bibinfo{author}{\bibfnamefont{E.}~\bibnamefont{Mart{\'\i}n-L{\'o}pez}},
  \bibinfo{author}{\bibfnamefont{N.~J.} \bibnamefont{Russell}},
  \bibinfo{author}{\bibfnamefont{J.~W.} \bibnamefont{Silverstone}},
  \bibinfo{author}{\bibfnamefont{P.~J.} \bibnamefont{Shadbolt}},
  \bibinfo{author}{\bibfnamefont{N.}~\bibnamefont{Matsuda}},
  \bibinfo{author}{\bibfnamefont{M.}~\bibnamefont{Oguma}},
  \bibinfo{author}{\bibfnamefont{M.}~\bibnamefont{Itoh}}, \bibnamefont{et~al.},
  \bibinfo{journal}{Science} \textbf{\bibinfo{volume}{349}},
  \bibinfo{pages}{711} (\bibinfo{year}{2015}).

\bibitem[{\citenamefont{Ladd et~al.}(2010)\citenamefont{Ladd, Jelezko,
  Laflamme, Nakamura, Monroe, and O’Brien}}]{Ladd2010}
\bibinfo{author}{\bibfnamefont{T.~D.} \bibnamefont{Ladd}},
  \bibinfo{author}{\bibfnamefont{F.}~\bibnamefont{Jelezko}},
  \bibinfo{author}{\bibfnamefont{R.}~\bibnamefont{Laflamme}},
  \bibinfo{author}{\bibfnamefont{Y.}~\bibnamefont{Nakamura}},
  \bibinfo{author}{\bibfnamefont{C.}~\bibnamefont{Monroe}}, \bibnamefont{and}
  \bibinfo{author}{\bibfnamefont{J.~L.} \bibnamefont{O’Brien}},
  \bibinfo{journal}{Nature} \textbf{\bibinfo{volume}{464}}, \bibinfo{pages}{45}
  (\bibinfo{year}{2010}).

\bibitem[{\citenamefont{Pan et~al.}(2012)\citenamefont{Pan, Chen, Lu,
  Weinfurter, Zeilinger, and \ifmmode~\dot{Z}\else
  \.{Z}\fi{}ukowski}}]{Pan2012}
\bibinfo{author}{\bibfnamefont{J.-W.} \bibnamefont{Pan}},
  \bibinfo{author}{\bibfnamefont{Z.-B.} \bibnamefont{Chen}},
  \bibinfo{author}{\bibfnamefont{C.-Y.} \bibnamefont{Lu}},
  \bibinfo{author}{\bibfnamefont{H.}~\bibnamefont{Weinfurter}},
  \bibinfo{author}{\bibfnamefont{A.}~\bibnamefont{Zeilinger}},
  \bibnamefont{and}
  \bibinfo{author}{\bibfnamefont{M.}~\bibnamefont{\ifmmode~\dot{Z}\else
  \.{Z}\fi{}ukowski}}, \bibinfo{journal}{Rev. Mod. Phys.}
  \textbf{\bibinfo{volume}{84}}, \bibinfo{pages}{777} (\bibinfo{year}{2012}).

\bibitem[{\citenamefont{Bouwmeester et~al.}(1997)\citenamefont{Bouwmeester,
  Pan, Mattle, Eibl, Weinfurter, and Zeilinger}}]{Bouwmeester1997}
\bibinfo{author}{\bibfnamefont{D.}~\bibnamefont{Bouwmeester}},
  \bibinfo{author}{\bibfnamefont{J.-W.} \bibnamefont{Pan}},
  \bibinfo{author}{\bibfnamefont{K.}~\bibnamefont{Mattle}},
  \bibinfo{author}{\bibfnamefont{M.}~\bibnamefont{Eibl}},
  \bibinfo{author}{\bibfnamefont{H.}~\bibnamefont{Weinfurter}},
  \bibnamefont{and}
  \bibinfo{author}{\bibfnamefont{A.}~\bibnamefont{Zeilinger}},
  \bibinfo{journal}{Nature} \textbf{\bibinfo{volume}{390}},
  \bibinfo{pages}{575} (\bibinfo{year}{1997}).

\bibitem[{\citenamefont{Mattle et~al.}(1996)\citenamefont{Mattle, Weinfurter,
  Kwiat, and Zeilinger}}]{Mattle1996}
\bibinfo{author}{\bibfnamefont{K.}~\bibnamefont{Mattle}},
  \bibinfo{author}{\bibfnamefont{H.}~\bibnamefont{Weinfurter}},
  \bibinfo{author}{\bibfnamefont{P.~G.} \bibnamefont{Kwiat}}, \bibnamefont{and}
  \bibinfo{author}{\bibfnamefont{A.}~\bibnamefont{Zeilinger}},
  \bibinfo{journal}{Phys. Rev. Lett.} \textbf{\bibinfo{volume}{76}},
  \bibinfo{pages}{4656} (\bibinfo{year}{1996}).

\bibitem[{\citenamefont{Lo et~al.}(2014)\citenamefont{Lo, Curty, and
  Tamaki}}]{Lo2014}
\bibinfo{author}{\bibfnamefont{H.-K.} \bibnamefont{Lo}},
  \bibinfo{author}{\bibfnamefont{M.}~\bibnamefont{Curty}}, \bibnamefont{and}
  \bibinfo{author}{\bibfnamefont{K.}~\bibnamefont{Tamaki}},
  \bibinfo{journal}{Nature Photonics} \textbf{\bibinfo{volume}{8}},
  \bibinfo{pages}{595} (\bibinfo{year}{2014}).

\bibitem[{\citenamefont{Dowling}(2008)}]{Dowling2008}
\bibinfo{author}{\bibfnamefont{J.~P.} \bibnamefont{Dowling}},
  \bibinfo{journal}{Contemporary Physics} \textbf{\bibinfo{volume}{49}},
  \bibinfo{pages}{125} (\bibinfo{year}{2008}).

\bibitem[{\citenamefont{Lemos et~al.}(2014)\citenamefont{Lemos, Borish, Cole,
  Ramelow, Lapkiewicz, and Zeilinger}}]{Lemos2014}
\bibinfo{author}{\bibfnamefont{G.~B.} \bibnamefont{Lemos}},
  \bibinfo{author}{\bibfnamefont{V.}~\bibnamefont{Borish}},
  \bibinfo{author}{\bibfnamefont{G.~D.} \bibnamefont{Cole}},
  \bibinfo{author}{\bibfnamefont{S.}~\bibnamefont{Ramelow}},
  \bibinfo{author}{\bibfnamefont{R.}~\bibnamefont{Lapkiewicz}},
  \bibnamefont{and}
  \bibinfo{author}{\bibfnamefont{A.}~\bibnamefont{Zeilinger}},
  \bibinfo{journal}{Nature} \textbf{\bibinfo{volume}{512}},
  \bibinfo{pages}{409} (\bibinfo{year}{2014}).

\bibitem[{\citenamefont{Motes et~al.}(2014)\citenamefont{Motes, Gilchrist,
  Dowling, and Rohde}}]{Motes2014}
\bibinfo{author}{\bibfnamefont{K.~R.} \bibnamefont{Motes}},
  \bibinfo{author}{\bibfnamefont{A.}~\bibnamefont{Gilchrist}},
  \bibinfo{author}{\bibfnamefont{J.~P.} \bibnamefont{Dowling}},
  \bibnamefont{and} \bibinfo{author}{\bibfnamefont{P.~P.} \bibnamefont{Rohde}},
  \bibinfo{journal}{Phys. Rev. Lett.} \textbf{\bibinfo{volume}{113}},
  \bibinfo{pages}{120501} (\bibinfo{year}{2014}).

\bibitem[{\citenamefont{Aaronson and Arkhipov}(2011)}]{Aaronson2011}
\bibinfo{author}{\bibfnamefont{S.}~\bibnamefont{Aaronson}} \bibnamefont{and}
  \bibinfo{author}{\bibfnamefont{A.}~\bibnamefont{Arkhipov}}, in
  \emph{\bibinfo{booktitle}{Proceedings of the Forty-third Annual ACM Symposium
  on Theory of Computing}} (\bibinfo{publisher}{ACM}, \bibinfo{address}{New
  York, NY, USA}, \bibinfo{year}{2011}), STOC '11, pp.
  \bibinfo{pages}{333--342}, ISBN \bibinfo{isbn}{978-1-4503-0691-1}.

\bibitem[{\citenamefont{Spagnolo et~al.}(2014)\citenamefont{Spagnolo, Vitelli,
  Bentivegna, Brod, Crespi, Flamini, Giacomini, Milani, Ramponi, Mataloni
  et~al.}}]{Spagnolo2013b}
\bibinfo{author}{\bibfnamefont{N.}~\bibnamefont{Spagnolo}},
  \bibinfo{author}{\bibfnamefont{C.}~\bibnamefont{Vitelli}},
  \bibinfo{author}{\bibfnamefont{M.}~\bibnamefont{Bentivegna}},
  \bibinfo{author}{\bibfnamefont{D.~J.} \bibnamefont{Brod}},
  \bibinfo{author}{\bibfnamefont{A.}~\bibnamefont{Crespi}},
  \bibinfo{author}{\bibfnamefont{F.}~\bibnamefont{Flamini}},
  \bibinfo{author}{\bibfnamefont{S.}~\bibnamefont{Giacomini}},
  \bibinfo{author}{\bibfnamefont{G.}~\bibnamefont{Milani}},
  \bibinfo{author}{\bibfnamefont{R.}~\bibnamefont{Ramponi}},
  \bibinfo{author}{\bibfnamefont{P.}~\bibnamefont{Mataloni}},
  \bibnamefont{et~al.}, \bibinfo{journal}{Nature Photonics}
  \textbf{\bibinfo{volume}{8}}, \bibinfo{pages}{615} (\bibinfo{year}{2014}).

\bibitem[{\citenamefont{Hong et~al.}(1987)\citenamefont{Hong, Ou, and
  Mandel}}]{Hong1987}
\bibinfo{author}{\bibfnamefont{C.~K.} \bibnamefont{Hong}},
  \bibinfo{author}{\bibfnamefont{Z.~Y.} \bibnamefont{Ou}}, \bibnamefont{and}
  \bibinfo{author}{\bibfnamefont{L.}~\bibnamefont{Mandel}},
  \bibinfo{journal}{Phys. Rev. Lett.} \textbf{\bibinfo{volume}{59}},
  \bibinfo{pages}{2044} (\bibinfo{year}{1987}).

\bibitem[{\citenamefont{Shih and Alley}(1988)}]{Shih1988}
\bibinfo{author}{\bibfnamefont{Y.~H.} \bibnamefont{Shih}} \bibnamefont{and}
  \bibinfo{author}{\bibfnamefont{C.~O.} \bibnamefont{Alley}},
  \bibinfo{journal}{Phys. Rev. Lett.} \textbf{\bibinfo{volume}{61}},
  \bibinfo{pages}{2921} (\bibinfo{year}{1988}).

\bibitem[{\citenamefont{Legero et~al.}(2003)\citenamefont{Legero, Wilk, Kuhn,
  and Rempe}}]{Legero2003}
\bibinfo{author}{\bibfnamefont{T.}~\bibnamefont{Legero}},
  \bibinfo{author}{\bibfnamefont{T.}~\bibnamefont{Wilk}},
  \bibinfo{author}{\bibfnamefont{A.}~\bibnamefont{Kuhn}}, \bibnamefont{and}
  \bibinfo{author}{\bibfnamefont{G.}~\bibnamefont{Rempe}},
  \bibinfo{journal}{Applied Physics B} \textbf{\bibinfo{volume}{77}},
  \bibinfo{pages}{797} (\bibinfo{year}{2003}).

\bibitem[{\citenamefont{Legero et~al.}(2004)\citenamefont{Legero, Wilk,
  Hennrich, Rempe, and Kuhn}}]{Legero2004}
\bibinfo{author}{\bibfnamefont{T.}~\bibnamefont{Legero}},
  \bibinfo{author}{\bibfnamefont{T.}~\bibnamefont{Wilk}},
  \bibinfo{author}{\bibfnamefont{M.}~\bibnamefont{Hennrich}},
  \bibinfo{author}{\bibfnamefont{G.}~\bibnamefont{Rempe}}, \bibnamefont{and}
  \bibinfo{author}{\bibfnamefont{A.}~\bibnamefont{Kuhn}},
  \bibinfo{journal}{Phys. Rev. Lett.} \textbf{\bibinfo{volume}{93}},
  \bibinfo{pages}{070503} (\bibinfo{year}{2004}).

\bibitem[{\citenamefont{Zhao et~al.}(2014)\citenamefont{Zhao, Zhang, Yang,
  Sang, Jiang, Bao, and Pan}}]{Zhao2014}
\bibinfo{author}{\bibfnamefont{T.-M.} \bibnamefont{Zhao}},
  \bibinfo{author}{\bibfnamefont{H.}~\bibnamefont{Zhang}},
  \bibinfo{author}{\bibfnamefont{J.}~\bibnamefont{Yang}},
  \bibinfo{author}{\bibfnamefont{Z.-R.} \bibnamefont{Sang}},
  \bibinfo{author}{\bibfnamefont{X.}~\bibnamefont{Jiang}},
  \bibinfo{author}{\bibfnamefont{X.-H.} \bibnamefont{Bao}}, \bibnamefont{and}
  \bibinfo{author}{\bibfnamefont{J.-W.} \bibnamefont{Pan}},
  \bibinfo{journal}{Phys. Rev. Lett.} \textbf{\bibinfo{volume}{112}},
  \bibinfo{pages}{103602} (\bibinfo{year}{2014}).

\bibitem[{\citenamefont{Vittorini et~al.}(2014)\citenamefont{Vittorini, Hucul,
  Inlek, Crocker, and Monroe}}]{Vittorini2014}
\bibinfo{author}{\bibfnamefont{G.}~\bibnamefont{Vittorini}},
  \bibinfo{author}{\bibfnamefont{D.}~\bibnamefont{Hucul}},
  \bibinfo{author}{\bibfnamefont{I.~V.} \bibnamefont{Inlek}},
  \bibinfo{author}{\bibfnamefont{C.}~\bibnamefont{Crocker}}, \bibnamefont{and}
  \bibinfo{author}{\bibfnamefont{C.}~\bibnamefont{Monroe}},
  \bibinfo{journal}{Phys. Rev. A} \textbf{\bibinfo{volume}{90}},
  \bibinfo{pages}{040302} (\bibinfo{year}{2014}).

\bibitem[{\citenamefont{Tamma and Laibacher}(2015)}]{Tamma2015}
\bibinfo{author}{\bibfnamefont{V.}~\bibnamefont{Tamma}} \bibnamefont{and}
  \bibinfo{author}{\bibfnamefont{S.}~\bibnamefont{Laibacher}},
  \bibinfo{journal}{Phys. Rev. Lett.} \textbf{\bibinfo{volume}{114}},
  \bibinfo{pages}{243601} (\bibinfo{year}{2015}).

\bibitem[{\citenamefont{Laibacher and
  Tamma}(2018{\natexlab{a}})}]{Laibacher2018}
\bibinfo{author}{\bibfnamefont{S.}~\bibnamefont{Laibacher}} \bibnamefont{and}
  \bibinfo{author}{\bibfnamefont{V.}~\bibnamefont{Tamma}},
  \bibinfo{journal}{Phys. Rev. A} \textbf{\bibinfo{volume}{98}},
  \bibinfo{pages}{053829} (\bibinfo{year}{2018}{\natexlab{a}}).

\bibitem[{\citenamefont{Wang et~al.}(2018)\citenamefont{Wang, Jing, Sun, Yang,
  Yu, Tamma, Bao, and Pan}}]{Wang2018}
\bibinfo{author}{\bibfnamefont{X.-J.} \bibnamefont{Wang}},
  \bibinfo{author}{\bibfnamefont{B.}~\bibnamefont{Jing}},
  \bibinfo{author}{\bibfnamefont{P.-F.} \bibnamefont{Sun}},
  \bibinfo{author}{\bibfnamefont{C.-W.} \bibnamefont{Yang}},
  \bibinfo{author}{\bibfnamefont{Y.}~\bibnamefont{Yu}},
  \bibinfo{author}{\bibfnamefont{V.}~\bibnamefont{Tamma}},
  \bibinfo{author}{\bibfnamefont{X.-H.} \bibnamefont{Bao}}, \bibnamefont{and}
  \bibinfo{author}{\bibfnamefont{J.-W.} \bibnamefont{Pan}},
  \bibinfo{journal}{Phys. Rev. Lett.} \textbf{\bibinfo{volume}{121}},
  \bibinfo{pages}{080501} (\bibinfo{year}{2018}).

\bibitem[{\citenamefont{Gerrits et~al.}(2015)\citenamefont{Gerrits, Marsili,
  Verma, Shalm, Shaw, Mirin, and Nam}}]{Gerrits2015}
\bibinfo{author}{\bibfnamefont{T.}~\bibnamefont{Gerrits}},
  \bibinfo{author}{\bibfnamefont{F.}~\bibnamefont{Marsili}},
  \bibinfo{author}{\bibfnamefont{V.~B.} \bibnamefont{Verma}},
  \bibinfo{author}{\bibfnamefont{L.~K.} \bibnamefont{Shalm}},
  \bibinfo{author}{\bibfnamefont{M.}~\bibnamefont{Shaw}},
  \bibinfo{author}{\bibfnamefont{R.~P.} \bibnamefont{Mirin}}, \bibnamefont{and}
  \bibinfo{author}{\bibfnamefont{S.~W.} \bibnamefont{Nam}},
  \bibinfo{journal}{Phys. Rev. A} \textbf{\bibinfo{volume}{91}},
  \bibinfo{pages}{013830} (\bibinfo{year}{2015}).

\bibitem[{\citenamefont{Jin et~al.}(2015)\citenamefont{Jin, Gerrits, Fujiwara,
  Wakabayashi, Yamashita, Miki, Terai, Shimizu, Takeoka, and Sasaki}}]{Jin2015}
\bibinfo{author}{\bibfnamefont{R.-B.} \bibnamefont{Jin}},
  \bibinfo{author}{\bibfnamefont{T.}~\bibnamefont{Gerrits}},
  \bibinfo{author}{\bibfnamefont{M.}~\bibnamefont{Fujiwara}},
  \bibinfo{author}{\bibfnamefont{R.}~\bibnamefont{Wakabayashi}},
  \bibinfo{author}{\bibfnamefont{T.}~\bibnamefont{Yamashita}},
  \bibinfo{author}{\bibfnamefont{S.}~\bibnamefont{Miki}},
  \bibinfo{author}{\bibfnamefont{H.}~\bibnamefont{Terai}},
  \bibinfo{author}{\bibfnamefont{R.}~\bibnamefont{Shimizu}},
  \bibinfo{author}{\bibfnamefont{M.}~\bibnamefont{Takeoka}}, \bibnamefont{and}
  \bibinfo{author}{\bibfnamefont{M.}~\bibnamefont{Sasaki}},
  \bibinfo{journal}{Opt. Express} \textbf{\bibinfo{volume}{23}},
  \bibinfo{pages}{28836} (\bibinfo{year}{2015}).

\bibitem[{\citenamefont{Laibacher and Tamma}(2015)}]{Laibacher2015}
\bibinfo{author}{\bibfnamefont{S.}~\bibnamefont{Laibacher}} \bibnamefont{and}
  \bibinfo{author}{\bibfnamefont{V.}~\bibnamefont{Tamma}},
  \bibinfo{journal}{Phys. Rev. Lett.} \textbf{\bibinfo{volume}{115}},
  \bibinfo{pages}{243605} (\bibinfo{year}{2015}).

\bibitem[{\citenamefont{Tamma and Laibacher}(2016{\natexlab{a}})}]{Tamma2016b}
\bibinfo{author}{\bibfnamefont{V.}~\bibnamefont{Tamma}} \bibnamefont{and}
  \bibinfo{author}{\bibfnamefont{S.}~\bibnamefont{Laibacher}},
  \bibinfo{journal}{Quantum Information Processing}
  \textbf{\bibinfo{volume}{15}}, \bibinfo{pages}{1241}
  (\bibinfo{year}{2016}{\natexlab{a}}).

\bibitem[{\citenamefont{Tamma and Laibacher}(2016{\natexlab{b}})}]{Tamma2016}
\bibinfo{author}{\bibfnamefont{V.}~\bibnamefont{Tamma}} \bibnamefont{and}
  \bibinfo{author}{\bibfnamefont{S.}~\bibnamefont{Laibacher}},
  \bibinfo{journal}{Journal of Modern Optics} \textbf{\bibinfo{volume}{63}},
  \bibinfo{pages}{41} (\bibinfo{year}{2016}{\natexlab{b}}).

\bibitem[{\citenamefont{Laibacher and
  Tamma}(2018{\natexlab{b}})}]{Laibacher2018b}
\bibinfo{author}{\bibfnamefont{S.}~\bibnamefont{Laibacher}} \bibnamefont{and}
  \bibinfo{author}{\bibfnamefont{V.}~\bibnamefont{Tamma}},
  \bibinfo{journal}{arXiv:1801.03832}  (\bibinfo{year}{2018}{\natexlab{b}}).

\bibitem[{\citenamefont{Pant and Englund}(2016)}]{Pant2016}
\bibinfo{author}{\bibfnamefont{M.}~\bibnamefont{Pant}} \bibnamefont{and}
  \bibinfo{author}{\bibfnamefont{D.}~\bibnamefont{Englund}},
  \bibinfo{journal}{Phys. Rev. A} \textbf{\bibinfo{volume}{93}},
  \bibinfo{pages}{043803} (\bibinfo{year}{2016}).

\bibitem[{\citenamefont{Chen et~al.}(2007)\citenamefont{Chen, Lee, and
  Kumar}}]{Chen2007}
\bibinfo{author}{\bibfnamefont{J.}~\bibnamefont{Chen}},
  \bibinfo{author}{\bibfnamefont{K.~F.} \bibnamefont{Lee}}, \bibnamefont{and}
  \bibinfo{author}{\bibfnamefont{P.}~\bibnamefont{Kumar}},
  \bibinfo{journal}{Phys. Rev. A} \textbf{\bibinfo{volume}{76}},
  \bibinfo{pages}{031804} (\bibinfo{year}{2007}).

\bibitem[{\citenamefont{Avenhaus et~al.}(2009)\citenamefont{Avenhaus, Eckstein,
  Mosley, and Silberhorn}}]{Avenhaus2009}
\bibinfo{author}{\bibfnamefont{M.}~\bibnamefont{Avenhaus}},
  \bibinfo{author}{\bibfnamefont{A.}~\bibnamefont{Eckstein}},
  \bibinfo{author}{\bibfnamefont{P.~J.} \bibnamefont{Mosley}},
  \bibnamefont{and}
  \bibinfo{author}{\bibfnamefont{C.}~\bibnamefont{Silberhorn}},
  \bibinfo{journal}{Opt. Lett.} \textbf{\bibinfo{volume}{34}},
  \bibinfo{pages}{2873} (\bibinfo{year}{2009}).

\bibitem[{\citenamefont{Mittal et~al.}(2017)\citenamefont{Mittal, Orre,
  Restelli, Salem, Goldschmidt, and Hafezi}}]{Mittal2017}
\bibinfo{author}{\bibfnamefont{S.}~\bibnamefont{Mittal}},
  \bibinfo{author}{\bibfnamefont{V.~V.} \bibnamefont{Orre}},
  \bibinfo{author}{\bibfnamefont{A.}~\bibnamefont{Restelli}},
  \bibinfo{author}{\bibfnamefont{R.}~\bibnamefont{Salem}},
  \bibinfo{author}{\bibfnamefont{E.~A.} \bibnamefont{Goldschmidt}},
  \bibnamefont{and} \bibinfo{author}{\bibfnamefont{M.}~\bibnamefont{Hafezi}},
  \bibinfo{journal}{Phys. Rev. A} \textbf{\bibinfo{volume}{96}},
  \bibinfo{pages}{043807} (\bibinfo{year}{2017}).

\bibitem[{\citenamefont{Davis et~al.}(2017)\citenamefont{Davis, Saulnier,
  Karpi\'{n}ski, and Smith}}]{Davis2017}
\bibinfo{author}{\bibfnamefont{A.~O.~C.} \bibnamefont{Davis}},
  \bibinfo{author}{\bibfnamefont{P.~M.} \bibnamefont{Saulnier}},
  \bibinfo{author}{\bibfnamefont{M.}~\bibnamefont{Karpi\'{n}ski}},
  \bibnamefont{and} \bibinfo{author}{\bibfnamefont{B.~J.} \bibnamefont{Smith}},
  \bibinfo{journal}{Opt. Express} \textbf{\bibinfo{volume}{25}},
  \bibinfo{pages}{12804} (\bibinfo{year}{2017}).

\bibitem[{\citenamefont{Kuo et~al.}(2016)\citenamefont{Kuo, Gerrits, Verma, and
  Nam}}]{Kuo2016}
\bibinfo{author}{\bibfnamefont{P.~S.} \bibnamefont{Kuo}},
  \bibinfo{author}{\bibfnamefont{T.}~\bibnamefont{Gerrits}},
  \bibinfo{author}{\bibfnamefont{V.~B.} \bibnamefont{Verma}}, \bibnamefont{and}
  \bibinfo{author}{\bibfnamefont{S.~W.} \bibnamefont{Nam}},
  \bibinfo{journal}{Opt. Lett.} \textbf{\bibinfo{volume}{41}},
  \bibinfo{pages}{5074} (\bibinfo{year}{2016}).

\bibitem[{\citenamefont{Slussarenko et~al.}(2017)\citenamefont{Slussarenko,
  Weston, Chrzanowski, Shalm, Verma, Nam, and Pryde}}]{Slussarenko2017}
\bibinfo{author}{\bibfnamefont{S.}~\bibnamefont{Slussarenko}},
  \bibinfo{author}{\bibfnamefont{M.~M.} \bibnamefont{Weston}},
  \bibinfo{author}{\bibfnamefont{H.~M.} \bibnamefont{Chrzanowski}},
  \bibinfo{author}{\bibfnamefont{L.~K.} \bibnamefont{Shalm}},
  \bibinfo{author}{\bibfnamefont{V.~B.} \bibnamefont{Verma}},
  \bibinfo{author}{\bibfnamefont{S.~W.} \bibnamefont{Nam}}, \bibnamefont{and}
  \bibinfo{author}{\bibfnamefont{G.~J.} \bibnamefont{Pryde}},
  \bibinfo{journal}{Nature Photonics} \textbf{\bibinfo{volume}{11}},
  \bibinfo{pages}{700} (\bibinfo{year}{2017}).

\bibitem[{\citenamefont{Deshpande et~al.}(2018)\citenamefont{Deshpande,
  Fefferman, Tran, Foss-Feig, and Gorshkov}}]{Deshpande2018}
\bibinfo{author}{\bibfnamefont{A.}~\bibnamefont{Deshpande}},
  \bibinfo{author}{\bibfnamefont{B.}~\bibnamefont{Fefferman}},
  \bibinfo{author}{\bibfnamefont{M.~C.} \bibnamefont{Tran}},
  \bibinfo{author}{\bibfnamefont{M.}~\bibnamefont{Foss-Feig}},
  \bibnamefont{and} \bibinfo{author}{\bibfnamefont{A.~V.}
  \bibnamefont{Gorshkov}}, \bibinfo{journal}{Phys. Rev. Lett.}
  \textbf{\bibinfo{volume}{121}}, \bibinfo{pages}{030501}
  (\bibinfo{year}{2018}).

\bibitem[{\citenamefont{Oren et~al.}(2017)\citenamefont{Oren, Mutzafi, Eldar,
  and Segev}}]{Oren2017}
\bibinfo{author}{\bibfnamefont{D.}~\bibnamefont{Oren}},
  \bibinfo{author}{\bibfnamefont{M.}~\bibnamefont{Mutzafi}},
  \bibinfo{author}{\bibfnamefont{Y.~C.} \bibnamefont{Eldar}}, \bibnamefont{and}
  \bibinfo{author}{\bibfnamefont{M.}~\bibnamefont{Segev}},
  \bibinfo{journal}{Optica} \textbf{\bibinfo{volume}{4}}, \bibinfo{pages}{993}
  (\bibinfo{year}{2017}).

\bibitem[{\citenamefont{Wang et~al.}(2019)\citenamefont{Wang, Suchkov,
  Titchener, Szameit, and Sukhorukov}}]{Wang2019}
\bibinfo{author}{\bibfnamefont{K.}~\bibnamefont{Wang}},
  \bibinfo{author}{\bibfnamefont{S.~V.} \bibnamefont{Suchkov}},
  \bibinfo{author}{\bibfnamefont{J.~G.} \bibnamefont{Titchener}},
  \bibinfo{author}{\bibfnamefont{A.}~\bibnamefont{Szameit}}, \bibnamefont{and}
  \bibinfo{author}{\bibfnamefont{A.~A.} \bibnamefont{Sukhorukov}},
  \bibinfo{journal}{Optica} \textbf{\bibinfo{volume}{6}}, \bibinfo{pages}{41}
  (\bibinfo{year}{2019}).

\end{thebibliography}

\providecommand{\noopsort}[1]{}\providecommand{\singleletter}[1]{#1}%

\end{document}



\title{Supplemental Material: Interference of Temporally Distinguishable Photons Using Frequency-Resolved Detection\\}

\author{Venkata Vikram Orre}
\affiliation{Joint Quantum Institute,NIST/University of Maryland, College Park, Maryland 20742, USA}
\affiliation{Department of Electrical and Computer Engineering and IREAP, University of Maryland, College Park, Maryland 20742, USA}

\author{Elizabeth A. Goldschmidt}
\affiliation{Joint Quantum Institute,NIST/University of Maryland, College Park, Maryland 20742, USA}
\affiliation{U.S. Army Research Laboratory, Adelphi, MD 20783, USA}

\author{Abhinav~Deshpande}
\affiliation{Joint Quantum Institute,NIST/University of Maryland, College Park, Maryland 20742, USA}
\affiliation{Joint Center for Quantum Information and Computer Science, NIST/University of Maryland, College Park, Maryland 20742, USA}

\author{Alexey~V.~Gorshkov}
\affiliation{Joint Quantum Institute,NIST/University of Maryland, College Park, Maryland 20742, USA}
\affiliation{Joint Center for Quantum Information and Computer Science, NIST/University of Maryland, College Park, Maryland 20742, USA}

\author{Vincenzo Tamma}
\affiliation{School of Mathematics and Physics, University of Portsmouth, Portsmouth PO1 3QL, United Kingdom}
\affiliation{Institute of Cosmology and Gravitation, University of Portsmouth, Portsmouth PO1 3FX, United Kingdom}

\author{Mohammad Hafezi}
\affiliation{Joint Quantum Institute,NIST/University of Maryland, College Park, Maryland 20742, USA}
\affiliation{Department of Electrical and Computer Engineering and IREAP, University of Maryland, College Park, Maryland 20742, USA}
\affiliation{Department of Physics, University of Maryland, College Park, Maryland 20742, USA}

\author{Sunil Mittal}
\affiliation{Joint Quantum Institute,NIST/University of Maryland, College Park, Maryland 20742, USA}
\affiliation{Department of Electrical and Computer Engineering and IREAP, University of Maryland, College Park, Maryland 20742, USA}

\maketitle

\section{S1: Experimental Setup}
\begin{figure}[h]
\centering
\includegraphics[width=0.9\textwidth]{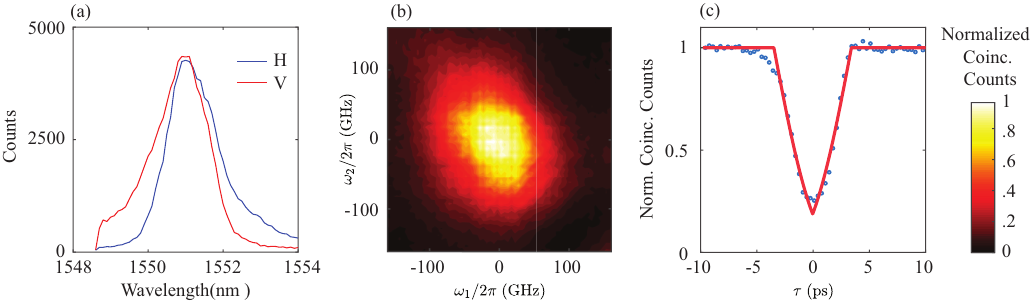}
\caption{\label{fig:S1}(a) Measured spectra of H and V polarized photons, (b) joint spectrum intensity (JSI) and, (c) HOM interference dip with a visibility of $82 \%$. The solid curve in (c) is an error function fit to the data \cite{Grice1997}.}
\end{figure}

\red We use type-II SPDC in a 30-mm-long PPKTP crystal to generate two- and four-photons in our experiment. The crystal has a poling period of $46.2~\mu\text{m}$ and is held at a temperature of $30^{\circ}~\text{C}$. We pump the crystal using a pulsed Ti:sapphire laser ($\approx$80 MHz, 1.6 ps) set at a wavelength of $\approx 775.5~\text{nm}$ such that the spectra of generated photons ($\approx 1551~\text{nm}$) is well within the passband of our CBG ($\approx$533 GHz, centered at 1550.9 nm). Figures~\ref{fig:S1}(a)-\ref{fig:S1}(c) show the spectra of the horizontally (H) and vertically (V) polarized photons measured using a tunable filter and a superconducting nanowire detector, the joint spectral intensity of the two-photon wavefunction measured using CBG as a time-of-flight spectrometer, and the HOM interference dip. We observe small ellipticity in the JSI that also manifests in the two-photon spectral correlations shown in Fig.2 of the main text. We also note that there is a small mismatch in the spectra of generated photons that reduces the visibility of the HOM interference. The visibility can be improved by spectral filtering or by using a crystal with extended phase matching \cite{Giovannetti2002}. Using HOM interference, we estimate the upper bound on the single-photon pulsewidth to be $\approx$1.7 ps [Fig.~\ref{fig:S1}(b)]. This agrees well the value of $\approx$1.55 ps that we estimate by fitting numerical simulation results to the experimentally measured data in the two-photon interference.

\begin{figure}
\centering
\includegraphics[width=0.6\textwidth]{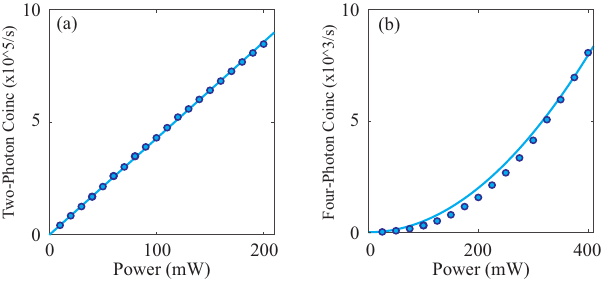}
\caption{\label{fig:S2}(a) Measured two-photon and, (b) four-photon generation rates as a function of pump power. The solid lines are linear and quadratic fits to the data, respectively.}
\end{figure}

The measured two- and four-photon generation rates as a function of pump power are shown in Fig.~\ref{fig:S2}. For the two-photon interference experiment (Fig. 2 of the main text), we used a pump power of 50 mW and the measured two-photon generation rate at the output of the SPDC was $\approx 2.1 \times 10^{5} ~\text{s}^{-1}$. For each delay setting, the data was acquired for 12 minutes and the maximum number of coincidence counts (normalization factor) in Figs. 2(a-d) is 313, 277, 267, and 198, respectively, in a 1 GHz frequency bin. For the three-photon interference experiment (Fig. 3 of the main text), we used a pump power of 400 mW and measured a four-photon generation rate of $\approx 8000 ~\text{s}^{-1}$. The data in Figs. 3(b)-3(e) was acquired for 270, 270, 180, and 390 minutes, and the maximum number of four-fold coincidence counts (normalization factor) is 325, 286, 289, and 167, respectively, in a 3 GHz frequency bin. For Fig. 4 measurements we pump the SPDC at 100 mW, and each data set was acquired for 20 minutes. The maximum number of counts in Figs. 4(a,b) is 180, in a frequency bin of 1 GHz. The insertion loss of the CBG was measured to be 1.7 dB, and the per-channel excess insertion losses of the fiber beam splitter and the tritter were measured to be $\approx$0.4 dB and $\approx$0.5 dB, respectively. \black

\section{S2: Fringe visibility in two-photon interference}

\begin{figure}[h]
\centering
\includegraphics[width=0.68\textwidth]{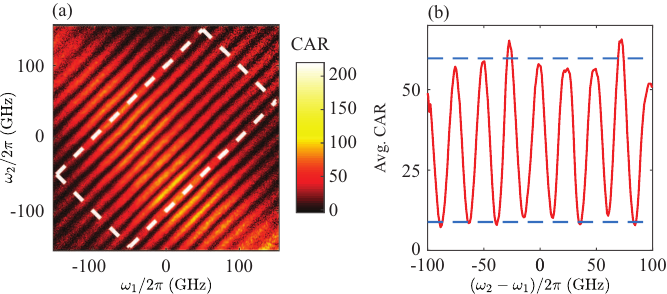}
\caption{\label{fig:S3} Measured CAR (a) as a function of $\W_{1}, \W_2$ and, (b) as a function of the frequency detuning $\left(\W_{2}-\W_{1} \right)$, for $\tau = 40 $ ps. Blue dotted lines show the mean maximum and minimum of the interference fringes.}
\end{figure}

Figure 2 of the main text shows interference fringes in the normalized coincidence counts as a function of frequency detuning $\left(\W_{2}-\W_{1}\right)$. From the plot, the interference fringe contrast seems to decrease with the increasing frequency detuning $\left(\W_{2}-\W_{1}\right)$. This apparent decrease in the contrast is mainly because of the lower coincidence counts as we move away from the center of the two-photon spectral intensity distribution. Therefore, to better estimate the visibility of the interference, we use the ratio of coincidence counts to accidental counts (product of singles counts on the two detectors), shown in Fig.~\ref{fig:S3}(a). We choose a window of 2 ns on each side from the center to further remove the contribution from the edges. Fig.~\ref{fig:S2}(b) shows CAR as a function of  $\left(\W_{2}-\W_{1} \right)$. From this plot, we calculate the fringe visibility as $\left(\overline{I_{\text{max}}} -\overline{I_{\text{min}}}\right)/\left(\overline{I_{\text{max}}}+\overline{I_{\text{min}}}\right)$, where $\overline{I_{\text{max}}}$ and $\overline{I_{\text{min}}}$ are averages of the peak maxima and minima, respectively. \red We used this visibility measurement scheme for all the data presented in Fig. 2 of the main text. \black

\section{S3: Miscellaneous probability contributions in three-photon interference}
\begin{figure}
 \centering
\includegraphics[width=0.7\textwidth]{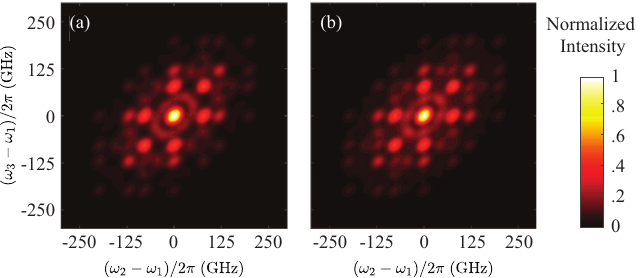}
\caption{\label{fig:S4} Simulated three-photon interference, (a) without and, (b) with the extra probability terms, for delays $\left(\tau_{21}, \tau_{31}\right) =  \left(10, 25\right)~\text{ps}$.}
\end{figure}

Our three-photon interference setup uses one polarization beam splitter and two non-polarizing beam splitters to probabilistically split two pairs of orthogonally polarized photons into four spatial modes (Fig. 3 of the main text). We use one of the modes to trigger the TIA, and use the other three modes along with delay lines and a tritter to achieve a configuration of three photons at times $\left(t_1, t_2, t_3\right)$ in one output port of the tritter (before the CBG). \red Due to the probabilistic nature of beam splitters, three photons with time-delays $\left(t_1, t_1, t_3\right)$ and  $\left(t_2, t_2, t_3\right)$ are also generated with a probability of 0.25, in addition to the desired $\left(t_1, t_2, t_3\right)$ events which occur with a probability of 0.5. \black In Fig.~\ref{fig:S4} we compare the simulated three-photon interference results, with and without these unwanted events. As mentioned in the main text, these events do not generate any new beat frequencies and therefore, do not drastically alter the three-photon interference pattern; we observe a fidelity of $\approx$0.99 between the two plots.

\section{S4: Reduced three-photon interference}
Here, we show how the integration of the three photon interference over one of the frequencies, reveals information about multiple pairwise interferences. The three-photon correlation function, $\Gamma\left(\W_{1},\W_{2},\W_{3} \right) = \psi^{*}\left(\W_{1},\W_{2},\W_{3}\right) \psi\left(\W_{1},\W_{2},\W_{3}\right)$ where, $\psi\left(\W_{1},\W_{2},\W_{3}\right)$ is the three-photon spectral wave function at the detectors, and is associated with the different ways in which three photons with spectral amplitudes $E_1$, $E_2$, $E_3$ at the input,injected at times $t_1$,$t_2$,$t_3$ can trigger the three detectors at frequencies $\W_{1}$,$\W_{2}$,$\W_{3}$. It is given as,
\begin{eqnarray}
\label{Eq1}
\nonumber \psi\left(\W_{1},\W_{2},\W_{3}\right)   &=& E_1\left(\W_{1}\right)E_2\left(\W_{2}\right)E_3\left(\W_{3}\right)e^{-i\W_1t_1}e^{-i\W_2t_2}e^{-i\W_3t_3} + \nonumber E_1\left(\W_{2}\right)E_2\left(\W_{1}\right)E_3\left(\W_{3}\right)e^{-i\W_2t_1}e^{-i\W_1t_2}e^{-i\W_3t_3} \\
\nonumber &+& E_1\left(\W_{1}\right)E_2\left(\W_{3}\right)E_3\left(\W_{2}\right)e^{-i\W_1t_1}e^{-i\W_3t_2}e^{-i\W_2t_3}
 + E_1\left(\W_{2}\right)E_2\left(\W_{3}\right)E_3\left(\W_{1}\right)e^{-i\W_2t_1}e^{-i\W_3t_2}e^{-i\W_1t_3} \\
 &+& E_1\left(\W_{3}\right)E_2\left(\W_{1}\right)E_3\left(\W_{2}\right)e^{-i\W_3t_1}e^{-i\W_1t_2}e^{-i\W_2t_3}
 + E_1\left(\W_{3}\right)E_2\left(\W_{2}\right)E_3\left(\W_{1}\right)e^{-i\W_3t_1}e^{-i\W_2t_2}e^{-i\W_1t_3}
\end{eqnarray}

\begin{figure}
 \centering
\includegraphics[width=0.98\textwidth]{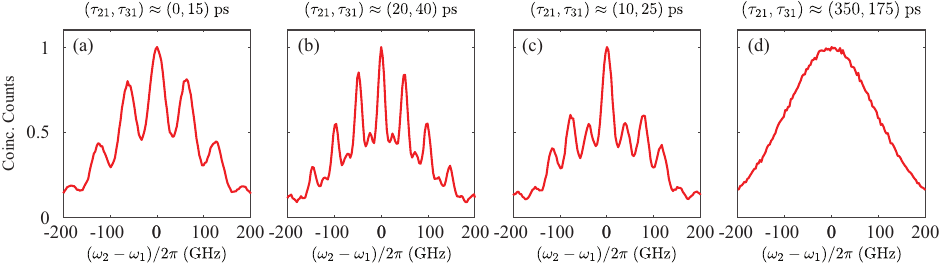}
\caption{\label{fig:S5} Measured three-photon correlation function integrated over $\W_{3}$, as a function of the frequency separation $\left(\W_{2}-\W_{1}\right)$ for different delay settings.}
\end{figure}

We rewrite the above equation by separating the contribution of $\W_3$ as,
\begin{equation}
\psi\left(\W_{1},\W_{2},\W_{3}\right) = \psi_{12}E_3\left(\W_{3}\right)e^{-i\W_3t_3} + \psi_{13}E_2\left(\W_{3}\right)e^{-i\W_3t_2} + \psi_{23}E_1\left(\W_{3}\right)e^{-i\W_3t_1}.
\end{equation}
Here
\begin{eqnarray}
\psi_{ij} = E_i\left(\W_{1}\right)e^{-i\W_1t_i}E_j\left(\W_{2}\right)e^{-i\W_2t_j} + E_j\left(\W_{1}\right)e^{-i\W_1t_j}E_i\left(\W_{2}\right)e^{-i\W_2t_i}
\end{eqnarray}
such that $\Gamma_{ij} = \left|\psi^{*}_{ij}\psi_{ij}\right|$ is the two-photon correlation function between photons with input times $t_i$ and $t_j$ respectively and is independent of $\W_3$. When the three-photon correlation function is integrated over measured frequencies at one of the detectors, here $\W_3$,
\begin{eqnarray}
\nonumber \int_{-\infty}^{\infty} \Gamma\left(\W_{1},\W_{2},\W_{3} \right) d\W_3 &=& \int_{-\infty}^{\infty} \psi^{*}\left(\W_{1},\W_{2},\W_{3}\right) \psi\left(\W_{1},\W_{2},\W_{3}\right) d\W_3 \\
\nonumber &=& \int_{-\infty}^{\infty} \Bigl[\left|\psi_{12}\right|^2\left|E_3\left(\W_{3}\right)\right|^2 + \left|\psi_{13}\right|^2\left|E_2\left(\W_{3}\right)\right|^2 + \left|\psi_{23}\right|^2\left|E_1\left(\W_{3}\right)\right|^2 \\
\nonumber &+& \psi^{*}_{12}\psi_{13}E_3^*\left(\W_{3}\right)E_2\left(\W_{3}\right)e^{-i\W_3(t_2-t_3)} + \psi^{*}_{12}\psi_{23}E_3^*\left(\W_{3}\right)E_1\left(\W_{3}\right)e^{-i\W_3(t_1-t_3)} \\
&+& \psi^{*}_{13}\psi_{23}E_2^*\left(\W_{3}\right)E_1\left(\W_{3}\right)e^{-i\W_3(t_1-t_2)} + \text{c.c} \Bigr] d\W_3
\end{eqnarray}
The first three terms are pairwise interferences between photon pairs (1,2), (1,3), (2,3). The remaining terms integrate out to zero, given the single photon spectral wave functions are smooth functions of $\W$. Therefore, the three-photon spectral correlation function integrated over one of the frequencies is given as
\begin{equation}
\int_{-\infty}^{\infty} \Gamma\left(\W_{1},\W_{2},\W_{3} \right) d\W_3 \propto \left|\psi_{12}\right|^2\left|E_3\left(\W_{3}\right)\right|^2 + \left|\psi_{13}\right|^2\left|E_2\left(\W_{3}\right)\right|^2 + \left|\psi_{23}\right|^2\left|E_1\left(\W_{3}\right)\right|^2,
\end{equation}
where, the intensities of three single-photon spectral wave functions contribute to the weights of the pairwise interferences. Fig.~\ref{fig:S5} shows the measured three-photon correlation function integrated over $\W_{3}$ and plotted as a function of the frequency separation $\left(\W_{2}-\W_{1}\right)$, for different delay settings. The Fourier transform of these spectra [Figs. 3(j)-3(m) of the main text] show the beat notes corresponding to the multiple pairwise interference terms between the three photons.

\section{S5: Generation of polarization/time-bin entangled photon pairs}

\begin{figure}[!h]
 \centering
\includegraphics[width=0.98\textwidth]{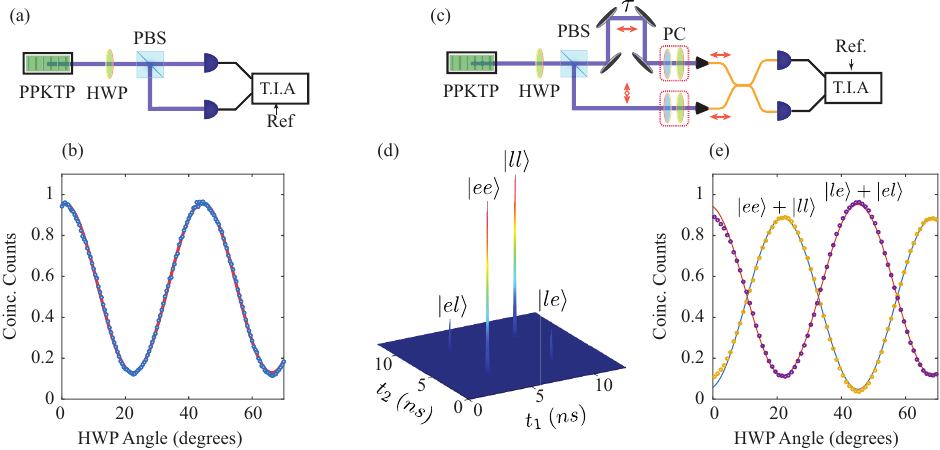}
\caption{\label{fig:S6}(a) Experimental setup to verify polarization entanglement. (b) Measured coincidences (blue) as a function of the half-wave plate angle. Red solid line is a cosine fit to the data. (c) Experimental setup to verify time-bin entanglement. (d) Measured joint-temporal intensity at HWP angle of $21.5^\circ$, and (e) coincidences as a function of the half-wave plate angle.}
\end{figure}

We use a HWP set to $22.5^\circ$ before the PBS of the setup shown in Fig. 1 to generate polarization-entangled entangled photon pairs in the two-photon interference setup [Fig.~\ref{fig:S6}(a)]. The HWP acts as a 50:50 beam splitter in the polarization domain and when the two-photon spectral wave function is symmetric under exchange of photons, generates a polarization-entangled state $\left|\Psi\right> = \left|2 \right>_{H} \left|0 \right>_{V} - \left|0 \right>_{H} \left|2 \right>_{V}$. To ensure the symmetric wave function, we use a band-pass filter with a bandwidth of $\approx 0.6$ nm. To verify this entanglement, we first measure the number coincidences in the two arms of the PBS as a function of the HWP angle \cite{Slussarenko2017} [Fig.~\ref{fig:S6}(b)]. The coincidence counts follow a cosine function with a period of $44(2)^\circ$, half of that expected for single photons. We observe the coincidences to be minimum at $22.5(5)^\circ$, that is, when the two photons are polarization-entangled. The interference visibility is calculated to be 83.0(6)$\%$. The low interference visibility is due to temporal walk-off between the H and V photons in the PPKTP crystal which leads to a small temporal distinguishability between the two photons at the HWP.

\red
We also verified the two-photon entanglement using coincidence measurements in the time domain [Fig.~\ref{fig:S6}(c)]. As before, we use a PBS to spatially separate the H and the V polarized photons. We then introduce a large delay ($\approx$5 ns) for  the V-polarized photons such that they can be temporally distinguished from the H-polarized photons. Subsequently, we change their polarization to H, and combine the two arms using a beam splitter. The H and the V polarization states now correspond to the `early'  and the `late' time-bin states, respectively. Fig.~\ref{fig:S6}(d) shows the measured temporal correlations between the two output ports of the beam splitter, when the HWP angle is set to $21.5^{\circ}$. As expected, we observe four coincidence peaks corresponding to the scenarios when both the photons arrive in the `early' time bin $\left(e-e\right)$, or in the `late' time bin $\left(l-l\right)$, and one photon arrives `early' and the other `late' $\left(e-l, ~\text{and} l-e \right)$. Fig.~\ref{fig:S6}(e) shows the total coincidence counts in the $e-e$ and $l-l$ bins, and in the $e-l$ and $l-e$ bins, as a function of the HWP angle. Similar to the polarization resolved measurements, we observe that the coincidence counts follow a cosine curve, and when the HWP angle is $\approx 22.5^\circ$, the two photons are in the time-bin entangled state  $\left|\Psi\right> = \left|2 \right>_{e} \left|0 \right>_{l} - e^{-i\varphi} \left|0 \right>_{e} \left|2 \right>_{l}$.  The interference visibility for coincidence counts in the $e-e$ and $l-l$ time bins is $\approx 90\%$ and that for the $e-l$ and $l-e$ time bins is $\approx 79\%$. \black

\section{S6: Stabilization of the interferometer}
\begin{figure}[!h]
 \centering
\includegraphics[width=0.8\textwidth]{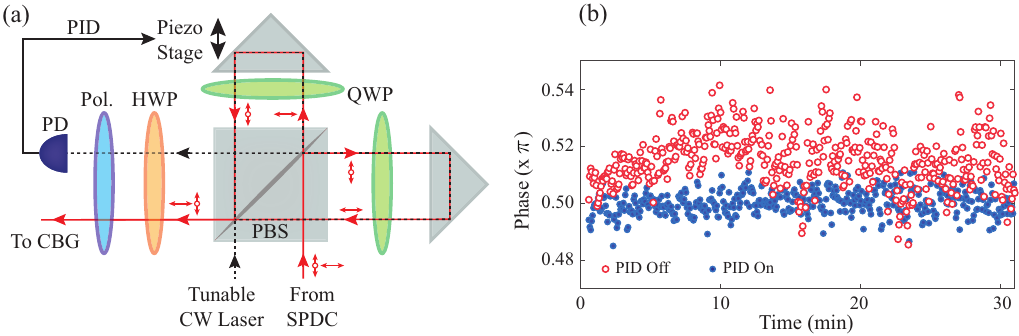}
\caption{\label{fig:S7}(a) Experimental setup to stabilize the interferometer. (b) Interferometer stability data with and without PID loop, when the interferometer phase was set to be $\pi/2$.}
\end{figure}

\red
For measurements using time-bin entangled photons (reported in Fig. 4 of the main text), we used a free-space interferometer [Fig.~\ref{fig:S7}(a)]. The interferometer was actively stabilized using a tunable CW laser and a PID feedback loop. The phase of the interferometer was changed by tuning the CW laser wavelength. During the course of the measurement, the interferometer phase was stable to within $\pm 1$ degree with the active stabilization and $\pm 2$ degrees without active feedback. \black

\section{S7: Exact sampling hardness in the multiphoton interference experiment}
\red
If our setup for the unentangled photons is scaled up to higher number of photons and modes, we sample from an output distribution described in \cite{Pant2016}. If the $j$'th input photon is emitted at time $t_j$ and gets detected at time $t$, the amplitude for this process is given by $U(t,t_j) = A(t,t_j) \star \mathcal{F}^{-1}\{e^{-i\phi(\omega) }\} $, where $A(t,t_j)$ is the shape of the input photon, the $\star$ operator corresponds to a Fourier convolution, $\mathcal{F}^{-1}$ denotes an inverse Fourier transformation, and $\phi(\omega)$ is the action of a (time-independent) dispersive element. We assume that the action of the dispersive element can be written as $\phi(\omega) = c_0 + c_1(\omega-\omega_0) + c_2(\omega-\omega_0)^2$. The nonlinear dispersion $c_2$ is what makes the photons spread out in time after passing through this dispersive element. Assuming the input photon wave packet has shape $A(t-t_j) \sim \exp[-\frac{(t-t_j)^2}{\Delta t^2}+i\omega_0 (t-t_j)]$, the convolution gives us an amplitude
\begin{align}
 U(t,t_j) &= \int_{-\infty}^\infty \mathrm{d}u A(u,t_j) \int_{-\infty}^\infty \mathrm{d}\omega \exp[-i\phi(\omega) + i\omega (t-u)] \\
 &\propto  \int_{-\infty}^\infty \mathrm{d}u  \int_{-\infty}^\infty \mathrm{d}\omega \exp\left[-\frac{(u-t_j)^2}{\Delta t^2} + i\omega_0 (u-t_j) -i\left(c_0 + c_1(\omega-\omega_0) + c_2(\omega-\omega_0)^2 \right) + i \omega (t-u)\right].
\end{align}
This simplifies to $U(t,t_j)= \psi(t-t_j) e^{i\omega_0t + 4ic_2 t^2}$, where
\begin{align}
 \psi(t-t_j) \propto \exp\left[ -\frac{(t-t_j-c_1)^2}{\Delta t^4 + 16c_2^2}(\Delta t^2 - 4ic_2)\right].
\end{align}
This amplitude $U(t,t_j)$ is also the expression for the matrix element of the linear-optical unitary once the temporal modes have been discretized by integrating over a small enough window of size $t_s$: $a^\dag_{t_j} = \frac{1}{\sqrt{t_s}}\int_{t_j}^{t_j+t_s} a^\dag(t) \mathrm{d}t$.

Now, when the input has photons emitted at times $\{0,\delta t, 2\delta t, \ldots\}$, and the detections at the output occur at times $\{t_1,t_2,\ldots\}$, the total amplitude for the output state is given by
\begin{align}
 \mathrm{Per}
 \begin{bmatrix}
  \psi(t_1) e^{i\omega_0t_1 + 4ic_2 t_1^2} & \psi(t_1-\delta t) e^{i\omega_0(t_1 -\delta t)+ 4ic_2 (t_1-\delta t)^2} & \psi(t_1-2\delta t) e^{i\omega_0(t_1 -2\delta t)+ 4ic_2 (t_1-2\delta t)^2} & \ldots \\
  \psi(t_2) e^{i\omega_0t_2 + 4ic_2 t_2^2} & \psi(t_2-\delta t) e^{i\omega_0(t_2 -\delta t)+ 4ic_2 (t_2-\delta t)^2} & \psi(t_2-2\delta t) e^{i\omega_0(t_2 -2\delta t)+ 4ic_2 (t_2-2\delta t)^2} & \ldots \\
  \vdots & \vdots & \ddots & \ldots
 \end{bmatrix}, \label{eq_unitary}
\end{align}
where the symbol ``$\mathrm{Per}$'' denotes the permanent of a matrix. In the two-photon case, one may verify that this gives fringes in the output distribution over $(t_1,t_2)$ with period $2\pi/(8c_2\delta t)$, which translates to the fringes in the measured frequencies $(\omega_1,\omega_2)$.

We now analyze the complexity of sampling from the output distribution for the $n$-photon case. The matrix in Eq.~\ref{eq_unitary} may be viewed as a submatrix of a bigger unitary matrix that has encoded in it all possible photon emission times and all possible detection times. This bigger unitary matrix has several symmetries since its columns differ only by translation in time and a phase. Further, it is approximately banded since the matrix elements fall off as a Gaussian.

The output distribution can be interpolated from being completely distinguishable ($ c_2 \rightarrow 0$) to indistinguishable ($ c_2^2 \gg \delta t$). In the indistinguishable regime, the output distribution can be hard to exactly sample from. Consider an event where the output photons are all detected at similar times, i.e.\ they are all within a small window around $c_1$. This event is unlikely since the photons that come later on have an exponentially suppressed amplitude for being detected at time $c_1$: $|U(c_1,k\delta t)| \sim \exp[-\frac{k^2 \delta t^2}{\Delta t^2 + 16c_2^2/\Delta t^2}] $, but the unlikeliness of such an outcome does not hamper a proof of exact sampling hardness.

The output amplitude for such an event is now given by the permanent of a matrix where the matrix elements $U_{ij}$ behave as $|U_{ij}| \sim \exp[-j^2/w]$, with $w = (\Delta t^2 + 16c_2^2/\Delta t^2)/\delta t^2$. The dependence on the row index $i$ is weak since all the output times were chosen to be around $t_1$. If one factors out this dependence of the columns to define a new matrix $R_{ij} = U_{ij}/\exp[-j^2/w]$, then one gets a matrix whose entries have uniform absolute values on average. It is plausible that randomly choosing the output time labels $(t_1,t_2,\ldots) = (c_1 + x_1, c_1 + x_2,\ldots )$ where $x_i$'s are random variables would give a sufficiently random matrix $R$. Aaronson and Arkhipov \cite{Aaronson2011} gave evidence that finding the permanent of a random matrix whose entries are random Gaussians is \#P-hard on average. The permanent of $U$ is simply $\exp[-n/w\sum_j j^2] \mathrm{Per\ } R$, which is a known factor. Hence $\mathrm{Per\ } U$ is as hard to compute as $\mathrm{Per\ } R$, which is likely \#P-hard on average. However, this does not rule out the possibility of an efficient approximate sampling algorithm that exploits the Gaussian fall-off in the matrix elements of the unitary.

\black

\providecommand{\noopsort}[1]{}\providecommand{\singleletter}[1]{#1}%